\documentclass[12pt]{article}
\usepackage{amsfonts}
\usepackage{amssymb}
\usepackage{graphics,amsmath}

\def\hybrid{\topmargin -20pt    \oddsidemargin 0pt
        \headheight 0pt \headsep 0pt
        \textwidth 6.35in       
        \textheight 9.25in       
        \marginparwidth .875in
        \parskip 5pt plus 1pt   \jot = 1.5ex}

\hybrid

\def\baselinestretch{1.2}

\catcode`\@=11

\def\marginnote#1{}
\newcount\hour
\newcount\minute
\newtoks\amorpm
\hour=\time\divide\hour by60
\minute=\time{\multiply\hour by60 \global\advance\minute by-\hour}
\edef\standardtime{{\ifnum\hour<12 \global\amorpm={am}%
        \else\global\amorpm={pm}\advance\hour by-12 \fi
        \ifnum\hour=0 \hour=12 \fi
        \number\hour:\ifnum\minute<10 0\fi\number\minute\the\amorpm}}
\edef\militarytime{\number\hour:\ifnum\minute<10 0\fi\number\minute}

\def\draftlabel#1{{\@bsphack\if@filesw {\let\thepage\relax
   \xdef\@gtempa{\write\@auxout{\string
      \newlabel{#1}{{\@currentlabel}{\thepage}}}}}\@gtempa
   \if@nobreak \ifvmode\nobreak\fi\fi\fi\@esphack}
        \gdef\@eqnlabel{#1}}
\def\@eqnlabel{}
\def\@vacuum{}
\def\draftmarginnote#1{\marginpar{\raggedright\scriptsize\tt#1}}

\def\draft{\oddsidemargin -.5truein
        \def\@oddfoot{\sl preliminary draft \hfil
        \rm\thepage\hfil\sl\today\quad\militarytime}
        \let\@evenfoot\@oddfoot \overfullrule 3pt
        \let\label=\draftlabel
        \let\marginnote=\draftmarginnote
   \def\@eqnnum{(\theequation)\rlap{\kern\marginparsep\tt\@eqnlabel}%
\global\let\@eqnlabel\@vacuum}  }

\def\preprint{\twocolumn\sloppy\flushbottom\parindent 2em
        \leftmargini 2em\leftmarginv .5em\leftmarginvi .5em
        \oddsidemargin -.5in    \evensidemargin -.5in
        \columnsep .4in \footheight 0pt
        \textwidth 10.in        \topmargin  -.4in
        \headheight 12pt \topskip .4in
        \textheight 6.9in \footskip 0pt
        \def\@oddhead{\thepage\hfil\addtocounter{page}{1}\thepage}
        \let\@evenhead\@oddhead \def\@oddfoot{} \def\@evenfoot{} }

\def\numberbysection{\@addtoreset{equation}{section}
        \def\theequation{\thesection.\arabic{equation}}}

\def\underline#1{\relax\ifmmode\@@underline#1\else
        $\@@underline{\hbox{#1}}$\relax\fi}

\def\titlepage{\@restonecolfalse\if@twocolumn\@restonecoltrue\onecolumn
     \else \newpage \fi \thispagestyle{empty}\c@page\z@
        \def\thefootnote{\fnsymbol{footnote}} }

\def\endtitlepage{\if@restonecol\twocolumn \else \newpage \fi
        \def\thefootnote{\arabic{footnote}}
        \setcounter{footnote}{0}}  

\catcode`@=12
\relax

\def\figcap{\section*{Figure Captions\markboth
        {FIGURECAPTIONS}{FIGURECAPTIONS}}\list
        {Figure \arabic{enumi}:\hfill}{\settowidth\labelwidth{Figure
999:}
        \leftmargin\labelwidth
        \advance\leftmargin\labelsep\usecounter{enumi}}}
 \relax
\def\tablecap{\section*{Table Captions\markboth
        {TABLECAPTIONS}{TABLECAPTIONS}}\list
        {Table \arabic{enumi}:\hfill}{\settowidth\labelwidth{Table
999:}
        \leftmargin\labelwidth
        \advance\leftmargin\labelsep\usecounter{enumi}}}
 \relax
\def\reflist{\section*{References\markboth
        {REFLIST}{REFLIST}}\list
        {[\arabic{enumi}]\hfill}{\settowidth\labelwidth{[999]}
        \leftmargin\labelwidth
        \advance\leftmargin\labelsep\usecounter{enumi}}}
 \relax
\makeatletter
\newcounter{pubctr}
\def\publist{\@ifnextchar[{\@publist}{\@@publist}}
\def\@publist[#1]{\list
        {[\arabic{pubctr}]\hfill}{\settowidth\labelwidth{[999]}
        \leftmargin\labelwidth
        \advance\leftmargin\labelsep
        \@nmbrlisttrue\def\@listctr{pubctr}
        \setcounter{pubctr}{#1}\addtocounter{pubctr}{-1}}}
\def\@@publist{\list
        {[\arabic{pubctr}]\hfill}{\settowidth\labelwidth{[999]}
        \leftmargin\labelwidth
        \advance\leftmargin\labelsep
        \@nmbrlisttrue\def\@listctr{pubctr}}}
 \relax
\makeatother
\newskip\humongous \humongous=0pt plus 1000pt minus 1000pt

\newif\ifdtup

\relax

\def\be{\begin{equation}}
\def\ee{\end{equation}}
\def\ba{\begin{eqnarray}}
\def\ea{\end{eqnarray}}

\def\a{\alpha}

\def\no{\noindent}

\def\IR{\relax{\rm I\kern-.18em R}}
\def\II{\relax{\rm 1\kern-.35em1}}

\renewcommand{\theequation}{\thesection.\arabic{equation}}
\csname @addtoreset\endcsname{equation}{section}

\def\IR{\relax{\rm I\kern-.18em R}}
\def\inv{^{\raise.15ex\hbox{${\scriptscriptstyle -}$}\kern-.05em 1}}

\begin{document}

\begin{titlepage}
\begin{center}

\vskip .5in

{\LARGE Elliptic solutions in the Neumann-Rosochatius system with mixed flux}
\vskip 0.4in

{\bf Rafael Hern\'andez} \phantom{x} and \phantom{x}
{\bf Juan Miguel Nieto} 
\vskip 0.1in

Departamento de F\'{\i}sica Te\'orica I \\
Universidad Complutense de Madrid \\
$28040$ Madrid, Spain \\
{\footnotesize{\tt rafael.hernandez@fis.ucm.es, juanieto@ucm.es}}

\end{center}

\vskip .4in

\centerline{\bf Abstract}
\vskip .1in
\no
Closed strings spinning in $AdS_3 \times S^3 \times T^4$ with mixed R-R and NS-NS three-form fluxes are described by a deformation of the 
one-dimensional Neumann-Rosochatius integrable system. In this article we find general solutions to this system that can be expressed in terms 
of elliptic functions. We consider closed strings rotating either in $S^3$ with two different angular momenta or in $AdS_3$ with one spin.  
In order to find the solutions we will need to extend the Uhlenbeck integrals of motion of the Neumann-Rosochatius system to include the contribution 
from the flux. In the limit of pure NS-NS flux, where the problem can be described by a supersymmetric WZW model, we find exact expressions 
for the classical energy in terms of the spin and the angular momenta of the spinning string. 

\noindent

\vskip .4in
\noindent

\end{titlepage}

\vfill
\eject

\def\baselinestretch{1.2}


\baselineskip 20pt


\section{Introduction}

The AdS$_3$/CFT$_2$ correspondence relates string theory backgrounds containing an $AdS_3$ factor and two-dimensional conformal field 
theories with maximal supersymmetry. As in other examples of the AdS/CFT correspondence, an integrable structure is also present in the case 
of AdS$_3$/CFT$_2$. The first explicit proof that integrability was a symmetry of the AdS$_3$/CFT$_2$ correspondence came from the observation 
that the Green-Schwarz action of type IIB string theory with R-R three-form flux compactified on $AdS_3 \times S^3 \times M_4$, with $M_4$ 
taken as either $T^4$ or $S^3 \times S^1$, is a classically integrable theory~\cite{BSZ}. This discovery lead immediately to an exhaustive analysis 
of various aspects of the AdS$_3$/CFT$_2$ correspondence using techniques inherited from other integrable systems~\cite{Zarembo}-\cite{Sundin} 
(see reference \cite{Sfondrini} for a comprehensive review). It was later on proved that integrability also remains a symmetry of more general string 
backgrounds including a mixture of R-R and NS-NS three-form fluxes \cite{CZ}. This result has been responsible for all the recent insight in the 
understanding of type IIB string theory on $AdS_3 \times S^3 \times T^4$ with mixed fluxes \cite{HT}-\cite{Stepanchuk}. 

Whenever integrability is present in a system, it shows up in many different facets. In the case of the AdS$_5$/CFT$_4$ correspondence an appealing 
approach to the search for the spectra of the theory came from the identification of the lagrangian describing closed strings rotating in $AdS_5 \times S^5$ 
with the Neumann-Rosochatius integrable system \cite{NR}. The Neumann-Rosochatius system is an integrable model describing an oscillator on a sphere 
or a hyperboloid with a centrifugal potential term. In reference \cite{HN} it was shown that the presence of non-vanishing NS-NS three-form flux introduces 
a deformation in the lagrangian of the Neumann-Rosochatius system. Integrability of the deformed system provides a systematic method to construct general 
solutions corresponding to closed string configurations rotating in $AdS_3 \times S^3 \times T^4$ with mixed R-R and NS-NS fluxes. The most immediate class 
of solutions that can be obtained in this way are the closed strings with constant radii found in \cite{HN} (string solutions with mixed fluxes were studied before using 
diverse approaches in references \cite{HST}-\cite{Babichenko}). The purpose of this article is to exploit the flux-deformed Neumann-Rosochatius system 
to construct more general solutions. 

The plan of the article is the following. In section~2 we will present the problem by considering an ansatz for a closed string rotating with two different 
angular momenta in~$S^3$ with NS-NS three-form flux. The resulting lagrangian is the Neumann-Rosochatius system with an additional contribution 
coming from the non-vanishing flux term. We will find the deformation introduced by the flux term in the Uhlenbeck integrals of motion of the system. 
In section~3 we will construct a general class of solutions with non-constant radii that can be expressed in terms of Jacobian elliptic functions. In the limit 
of pure NS-NS flux the problem can be described by a supersymmetric WZW model. In this limit the elliptic solutions reduce to trigonometric functions 
and we can find compact expressions for the classical energy of the rotating strings in terms of the angular momenta. In section~4 we extend the analysis 
to the case where the string is rotating in $AdS_{3}$. We conclude in section~5 with several remarks and some discussion on our results. We include 
an appendix where in the limit of pure NS-NS flux we solve the general case where the string is allowed to rotate both in $AdS_{3}$ and $S^3$.


\section{The Neumann-Rosochatius system with mixed flux}

In this article we will analyze the motion of closed strings rotating in $AdS_3 \times S^3 \times T^4$ with non-vanishing NS-NS three-form flux. 
We will consider no dynamics along the torus, and thus the background metric will be
\be
ds^2 = - \cosh^2 \rho \, dt^2 + d \rho^2 + \sinh^2 \rho \, d \phi^2 + 
d\theta^2 + \sin^2 \theta d\phi_1^2 + \cos^2 \theta d \phi_2^2 \ , 
\ee
and the NS-NS B-field will be 
\be
b_{t \phi} = q \sinh^2 \rho \ , \quad b_{\phi_1 \phi_2} = - q \cos^2 \theta \ , 
\label{Bfield}
\ee
where $0 \leq q \leq 1$. The limit $q=0$ corresponds to the case of pure R-R flux, while setting $q=1$ we are left with pure NS-NS flux. 
In the case of pure R-R flux the sigma model for closed strings rotating in $AdS_3 \times S^3$ can be reduced to the Neumann-Rosochatius 
integrable system \cite{NR}. The presence of NS-NS flux introduces an additional term in the Lagrangian of the Neumann-Rosochatius system \cite{HN}. 
In order to exhibit this it is convenient to use the embedding coordinates of $AdS_3$ and $S^3$, which are related to the global angles by
\begin{eqnarray}
Y_1 + i Y_2 \! \! \! & = & \! \! \! \sinh \rho \, e^{i \phi} \ , \quad Y_3 + i Y_0 = \cosh \rho \, e^{i  t} \ , \\
X_1 + i X_2 \! \! \! & = & \! \! \! \sin \theta \, e^{i \phi_1} \ , \quad X_3 + i X_4 = \cos \theta \, e^{i \phi_2} \ . 
\end{eqnarray}
For simplicity in this section and section 3 we will restrict the motion of the strings to rotation on $S^3$. The extension to strings spinning in $AdS_3$ 
will be considered in section 4. We will thus take $Y_3 + i Y_0 = e^{i w_0 \tau}$ and $Y_1=Y_2=0$, together with an ansatz for a closed string rotating 
with two different angular momenta along $S^3$, 
\be
X_1 + i X_2 = r_1 (\sigma) \, e^{i \varphi_1 (\tau, \sigma)} \ , \quad X_3 + i X_4 = r_2 (\sigma) \, e^{i \varphi_2 (\tau, \sigma)} \ .
\label{ansatz}
\ee
The angles are chosen as
\be
\varphi_i (\tau,\sigma) = \omega_i \tau + \alpha_i(\sigma) \ ,
\label{ansatzangle}
\ee
and the condition that the solutions lie on a three-sphere implies that 
\be
r_1^2 + r_2^2 = 1 \ . 
\label{threesphere}
\ee
When we enter this ansatz in the world-sheet action in the conformal gauge we find 
\be
L_{S^3} = \frac {\sqrt{\lambda}}{2 \pi} \Big[  \sum_{i=1}^2 \frac {1}{2} \big[ (r_i')^2 + r_i^2 (\a_i')^2 - r_i^2 \omega_i^2 \big] 
- \frac {\Lambda}{2} ( r_1^2 + r_2^2 - 1) + q r_2^2 \, ( \omega_1 \alpha_2' - \omega_2 \alpha_1' ) \Big] \ ,
\label{NRq}
\ee
where the prime stands for derivatives with respect to $\sigma$ and $\Lambda$ is a Lagrange multiplier. 
The equations of motion for the radial coordinates following from (\ref{NRq}) are
\begin{align}
r_1'' & = -r_1 \omega_1^2 +r_1 \alpha_1^{'2}-\Lambda r_1 \ , \label{r1prime} \\
r_2'' & = -r_2 \omega_2^2 +r_2 \alpha_2^{'2}-\Lambda r_2 + 2 q r_2 ( \omega_1 \alpha_2' - \omega_2 \alpha_1' ) \ . \label{r2prime}
\end{align}
The equations for the angles can be easily integrated once,
\be
\alpha_i' = \frac {v_i + q r_2^2 \epsilon_{ij} \omega_j}{r_i^2} \ , \quad i = 1,2 \ ,
\label{alphaprime}
\ee
where $v_i$ are some integration constants and we have chosen $\epsilon_{12} = +1$. These equations must be accompanied 
by the Virasoro constraints which read
\begin{align}
& \sum_{i=1}^2 \big[ r_i^{'2} + r_i^2 ( \alpha^{'2}_i + \omega_i^2 ) \big] = w_0^2 \ , \label{Virasoro1} \\
& \sum_{i=1}^2 r_i^2 \omega_i \alpha_i'= \sum_{i=1}^2 v_i \omega_i =0 \ . \label{Virasoro2} 
\end{align}
Furthermore dealing with closed string solutions requires 
\be
r_i (\sigma + 2 \pi) = r_i (\sigma) \ , \quad \alpha_i(\sigma + 2 \pi) = \alpha_i(\sigma) + 2 \pi \bar{m}_i \ ,
\label{periodicity}
\ee
where $\bar{m}_i$ are integer numbers that behave as winding numbers. The energy and the two angular momenta of the string are given by 
\begin{align}
E & = \sqrt{\lambda} \, w_0 \ , \label{E} \\
J_1 & = \sqrt{\lambda} \int_0^{2\pi} {\frac {d \sigma}{2 \pi} \left( r^2_1 \omega_1 - q r_2^2 \alpha_2' \right)} \ , \label{J1} \\
J_2 & = \sqrt{\lambda} \int_0^{2 \pi} {\frac {d \sigma}{2 \pi} \left( r^2_2 \omega_2 + q r_2^2 \alpha_1' \right)} \label{J2} \ .
\end{align}

Before concluding our presentation of the flux-deformed Neumann-Rosochatius Lagrangian 
we must note that there is a gauge freedom in the choice of the NS-NS two-form~(\ref{Bfield}). As a consequence of this ambiguity 
the flux term in~(\ref{NRq}) could have been written in the form 
$q \big[ (2-c) r_2^2 - c r_1^2 \big]/2$, where $c$ is an arbitrary parameter. 
This parameter does not affect the equations of motion, but it contributes as a total derivative to the angular momenta and thus it could show up 
if non-trivial boundary conditions were imposed.~\footnote{See reference \cite{HST} for a detailed discussion on this point concerning the dyonic giant magnon solutions.} 
However for the spinning string solutions that we will consider in this article the $c$-term plays no role because it can be absorbed in 
a redefinition of the $v_i$ integration constants and a shift of the angular momenta.  


\subsection{Integrals of motion}

The integrability of the Neumann-Rosochatius system follows from the existence of a set of integrals of motion in involution, 
the Uhlenbeck constants. The Uhlenbeck constants were first found in \cite{Uhlenbeck} for the case 
$\alpha_i= \hbox{constant}$ and were extended to general values of $\alpha_i$ in~\cite{NR}. 
In the case of a closed string rotating in $S^3$ there are two integrals $I_1$ and $I_2$, 
but as they must satisfy the constraint $I_1 + I_2 = 1$ we are left with a single independent constant,
\be
I_1 = r_1^2 + \frac{1}{\omega_1^2 - \omega_2^2} \left[ (r_1 r'_2 - r'_1 r_2)^2 
+ \frac{v_1^2}{r_1^2} r_2^2 + \frac{v_2^2}{r_2^2} r_1^2 \right] \ . 
\ee
Furthermore the Hamiltonian of the Neumann-Rosochatius system,
\be
H = \frac {1}{2} \sum_{i=1}^2 \big[ r_i^{\prime 2} +r_i^2 \alpha_i^{\prime 2} + r_i^2 \omega_i^2 \big] \ ,
\ee
can be written in terms of the Uhlenbeck constants and the integrals of motion $v_i$,  
\be
H = \frac {1}{2} \sum_{i=1}^2 \big[ \omega_i^2 I_i + v_i^2 \big] \ .
\ee
When the NS-NS three-form is turned on the Uhlenbeck constants should be deformed in some way. In order to find this deformation we will 
assume that the extended constant can be written as
\be
\bar{I}_1=r_1^2 +\frac{1}{\omega_1^2 -\omega_2^2} \left[ (r_1 r'_2-r'_1 r_2)^2 +\frac{v_1^2}{r_1^2} r_2^2 +\frac{v_2^2}{r_2^2} r_1^2 +2 f \right] \ ,  
\label{I1tilde}
\ee
where $f=f(r_1,r_2,q)$, with no dependence on $r'_1$ or $r'_2$. 
This function can be determined if we impose that $\bar{I}_1'=0$. After some immediate algebra we find that
\be
f' + \frac{(q^2 \omega_2^2 +2q\omega_2 v_1) r'_1}{r_1^3} +q^2 (\omega_1^2 -\omega_2^2) r_1 r'_1 =0 \ ,
\label{fprime}
\ee
where we have used the constraint (\ref{threesphere}) together with 
\be
r_1 r'_1 + r_2 r'_2 = 0 \ , \quad r_1 r''_1+(r'_1)^2 + r_2 r''_2 +(r'_2)^2 = 0 \ , 
\ee
and the equations of motion (\ref{r1prime}), (\ref{r2prime}) and (\ref{alphaprime}). As all three terms in relation (\ref{fprime}) are total derivatives, 
integration is immediate and we readily conclude that the deformation of the Uhlenbeck constant is given by
\be
\bar{I}_1 = r_1^2 (1-q^2) +\frac{1}{\omega_1^2 -\omega_2^2} \left[ (r_1 r'_2-r'_1 r_2)^2 + \frac{(v_1+q\omega_2)^2}{r_1^2} r_2^2 + \frac{v_2^2}{r_2^2} r_1^2 \right] \ .
\label{deformedUhlenbeck}
\ee
As in the absence of flux, the deformed constants satisfy the condition $\bar{I}_1+\bar{I}_2=1$. \footnote{Note that if we had included the parameter $c$ in the Lagrangian the 
Uhlenbeck integrals of motion would have satisfied also this constraint because $c$ can be absorbed in the $v_i$ integrals of motion.}
The Hamiltonian including the contribution from the NS-NS flux can also be written now using the deformed Uhlenbeck constants and the integrals of motion $v_i$,
\be
H = \frac {1}{2} \sum_{i=1}^2 \big[ \omega_i^2 \bar{I}_i + v_i^2 \big] + \frac {1}{2} q^2 (\omega_1^2 - \omega_2^2 ) - q \omega_1 v_2 \ .
\label{H}
\ee


\section{Spinning strings in $S^3$}

In this section we will construct general solutions of the Neumann-Rosochatius system in the presence on NS-NS flux corresponding to closed strings 
spinning in $S^3$ with two different angular momenta. A convenient way to analyze this problem is to introduce an ellipsoidal coordinate \cite{BT}. 
The ellipsoidal coordinate $\zeta$ is defined as the root of the equation
\be
\frac{r_1^ 2}{\zeta -\omega_1^ 2} + \frac{r_2^2}{\zeta -\omega_2^2} = 0 \ .
\ee
If we choose the angular frequencies such that $\omega_1 < \omega_2$ the range of the ellipsoidal coordinate is $\omega_1^2 \leq \zeta \leq \omega_2^2$. 
Now we could enter $\zeta$ directly into the equations of motion to find the second order differential equation for this coordinate. Instead we will follow \cite{NR} 
and use the Uhlenbeck constants to reduce the problem to a first order differential equation. In order to do this we just need to 
write the Uhlenbeck integral in terms of the ellipsoidal coordinate. This is immediate once we note that
\be
(r_1 r'_2 -r_2 r'_1)^2=\frac{\zeta^{\prime 2}}{4 (\omega_1^2 -\zeta) (\zeta -\omega_2^2)} \ .
\ee
When we solve for $\zeta'^2$ in the deformed Uhlenbeck constant (\ref{deformedUhlenbeck}) we conclude that
\be
\zeta'^2 = - 4 P_3(\zeta) \ ,
\label{diffeqzeta}
\ee
where $P_3(\zeta)$ is the third order polynomial 
\begin{align}
P_3(\zeta)
& = (1-q^2) (\zeta -\omega_1^2 )^2 (\zeta -\omega_2^2) + (\zeta- \omega_1^2) (\zeta- \omega_2^2) (\omega_1^2 -\omega_2^2) \bar{I}_1 \notag \\
& + (\zeta - \omega_1^2)^2  v_2^2 + (\zeta - \omega_2^2)^2 (v_1 + q \omega_2)^2 = (1-q^2 ) \prod_{i=1}^3 (\zeta -\zeta_i)  \ .
\end{align}
This polynomial defines an elliptic curve $s^2 + P_3(\zeta)=0$. In fact if we change variables to
\be
\zeta = \zeta_2 + (\zeta_3 - \zeta_2) \eta^2 \ ,
\ee
equation (\ref{diffeqzeta}) becomes the differential equation for the Jacobian elliptic cosine,
\be
\eta^{\prime 2}=(1-q^2) (\zeta_3 - \zeta_1) (1 - \eta ^2) (1- \kappa + \kappa \eta^2) \ ,
\ee
where the elliptic modulus is given by $\kappa = (\zeta_3 - \zeta_2)/(\zeta_3 - \zeta_1)$. The solution is thus
\be
\eta(\sigma) = \hbox {cn} \big( \sigma \sqrt{(1-q^2) (\zeta_3 -\zeta_1)} +\sigma_0, \kappa \big) \ ,
\label{etasn}
\ee
with $\sigma_0$ an integration constant that can be set to zero by performing a rotation. Therefore we conclude that
\be
r_1^2(\sigma) = \frac{\zeta_3 - \omega_1^2}{\omega_2^2 - \omega_1^2} + \frac{\zeta_2 - \zeta_3}{\omega_2^2 - \omega_1^2} \, \hbox{sn}^2
\big( \sigma \sqrt{(1-q^2) (\zeta_3 -\zeta_1)} , \kappa \big) \ .
\label{r1elliptic}
\ee
We must stress that we need to order the roots in such a way that $\zeta_1 < \zeta_3$ to make sure that the argument of the elliptic sine is real. We also need 
$\zeta_2 < \zeta_3$ to have $\kappa > 0$, together with $\zeta_1 < \zeta_2 $ to keep $\kappa < 1$. Furthermore, 
imposing that (\ref{r1elliptic}) must have codomain between $0$ and $1$ demands 
$\omega_1^2 \leq \zeta_{2,3} \leq \omega_2^2$. Note that this restriction does not apply to $\zeta_1$. This is a solution 
of circular type.~\footnote{In the absence of R-R flux and setting the $v_i$ integrals to zero, this choice of parameters 
corresponds to solutions of circular type when $I_1$ is taken as negative, or solutions of folded type when $I_1$ is positive~\cite{NR}. } 
The periodicity condition on the radial coordinates implies that~\footnote{There are four cases in which we have to alter this periodicity condition.
When either $v_1+ q \omega_2 = 0$ or $v_2 = 0$ the condition becomes $\frac{\pi}{2} \sqrt{(1-q^2) (\zeta_3 -\zeta_1)} = n \, \hbox{K} (\kappa )$ 
because of changes of branch when we take the square root of (\ref{r1elliptic}). The two remaining cases correspond to the limit $\zeta_3 \rightarrow \zeta_2$, 
which is the constant radii case, and to the limit $\kappa \rightarrow 1$, where the periodicity of the elliptic sine becomes infinite. In both cases 
there is no periodicity condition. We will discuss these two limits below in this section.}
\be
\pi \sqrt{(1-q^2) (\zeta_3 -\zeta_1)} = n \, \hbox{K} (\kappa ) \ , 
\ee
where we have used that $2\hbox{K}(\kappa)$ is the period of the square of the Jacobi sine, with 
$\hbox{K}(\kappa)$ the complete elliptic integral of the first kind and $n$ an integer number. We can use now equation (\ref{r1elliptic}) to write the winding numbers 
$\bar{m}_i$ in terms of the integration constants $v_i$ and the angular frequencies $\omega_i$. From the periodicity condition on $\alpha_1$,
\be
2 \pi \bar{m}_1 = \int_0^{2\pi}{\alpha'_1 d\sigma}=\int_0^{2\pi}{\left( \frac{v_1}{r_1^2} + q \omega_2 \frac{r_2^2}{r_1^2} \right) d\sigma} \ ,
\ee
we can write
\be
\frac{\bar{m}_1 + q \omega_2}{v_1 + q \omega_2}= \int_0^{2\pi} \frac{d\sigma}{2\pi} \frac{1}{r_1^2} \ .
\ee
Inserting (\ref{r1elliptic}) in this expression and performing the integration we find
\be
\bar{m}_1 + q \omega_2 = \frac{(v_1 + q \omega_2) (\omega_2^2 - \omega_1^2)}{(\zeta_3 - \omega_1^2)
\hbox{K} (\kappa)} \Pi \left( \frac{\zeta_3 - \zeta_2}{\zeta_3 - \omega_1^2} , \kappa \right) \ ,
\ee
where $\Pi(a,b)$ is the complete elliptic integral of the third kind. In a similar way, from the periodicity condition for $\alpha_2$, 
\be
2 \pi \bar{m}_2 = \int_0^{2\pi}{\alpha'_2 d\sigma} = \int_0^{2\pi}{\left( \frac{v_2}{r_2^2} - q \omega_1 \right) d \sigma} \ ,
\ee
we find that
\be
\frac{\bar{m}_2 + q \omega_1}{v_2}= \int_0^{2\pi} \frac{d\sigma}{2\pi} \frac{1}{r_2^2} \ ,
\ee
that we can integrate to get
\be
\bar{m}_2 + q \omega_1 = \frac{v_2 (\omega_2^2 - \omega_1^2 )}{(\omega_2^2 - \zeta_3) \hbox{K} (\kappa)} \Pi 
\left( - \ \frac{\zeta_3 - \zeta_2}{\omega_2^2 - \zeta_3} ,\kappa \right) \ .
\ee
We can perform an identical computation to obtain the angular momenta. From equation (\ref{J1}) we get
\be
\frac{J_1} {\sqrt{\lambda}} + q v_2 - q^2 \omega_1 = \omega_ 1 (1-q^2) \int_0^{2\pi}{\frac{d\sigma}{2\pi} r_1^2} \ ,
\ee
and therefore
\be
\frac{J_1}{\sqrt{\lambda}} = \frac{\omega_1 (1-q^2)}{\omega_2^2 -\omega_1^2} \left[ \zeta_3 - \omega_1^2 - (\zeta_3 - \zeta_1) \
\left( 1-\frac{ \hbox{E} (\kappa)}{\hbox{K} (\kappa)} \right) \right] - q v_2 +q^2 \omega_1 \ .
\label{J1elliptic}
\ee
with $\hbox{E}(\kappa)$ the complete elliptic integral of the second kind. As before, (\ref{J2}) implies
\be
\frac{J_2}{\sqrt{\lambda}} + q v_1 - q \bar{m}_1= \omega_2 (1-q^2) \int_0^{2\pi}{\frac{d\sigma}{2\pi} r_2^2} \ ,
\ee
and thus after integration we conclude that
\be
\frac {J_2}{\sqrt{\lambda}} = \frac{\omega_2 (1-q^2)}{\omega_2^2 -\omega_1^2} \left[ \omega_2^2 -\zeta_3 + (\zeta_3 - \zeta_1) \
\left( 1-\frac{\hbox{E}(\kappa)}{\hbox{K}(\kappa)} \right) \right] - q v_1 +q \bar{m}_1 \ .
\label{J2elliptic}
\ee
These expressions for the angular momenta can be used to rewrite the first Virasoro constrain (\ref{Virasoro2}) as
\be
\omega_2 J_1 +\omega_1 J_2 = \sqrt{\lambda} \left( \omega_1 \omega_2 +q\omega_1 \bar{m}_1 \right) \ .
\ee
In principle we could now employ these relations to write the energy in terms of the winding numbers $\bar{m}_i$, 
the angular momenta $J_i$ and the Uhlenbeck constants. The first step to achieve this is to express the energy 
in terms of the frequencies $\omega_i$ and the constants $v_i$. Then we need to write the $v_i$ in terms of the angular momenta 
using relations (\ref{J1elliptic}) and (\ref{J2elliptic}). However solving these equations for arbitrary values of $q$ is a very involved problem and we will not present a discussion on it here. 
Instead in the following subsection we will focus on the only case where a simple analysis is possible which is that of pure NS-NS flux. 


But before exploring our solutions in the regime of pure NS-NS flux we will discuss an important limit where they can be reduced to simpler ones. 
It corresponds to the choices of parameters that make the discriminant of $P_3(\zeta)$ equal to zero. Our hierarchy of roots implies that there are only three cases able to fulfill 
this condition. The first corresponds to solutions with constant radii, where $\zeta_2=\zeta_3$. These solutions were first constructed in~\cite{HST} 
and later on recovered by deriving the corresponding finite-gap equations in~\cite{Babichenko} or by solving the equations of motion for the flux-deformed 
Neumann-Rosochatius system in~\cite{HN}. The second case corresponds to the limit $\kappa=1$, which is obtained when $\zeta_1=\zeta_2$. 
These are the giant magnons analyzed in~\cite{BPP} for the $v_2=0$ case and in~\cite{ABozhilov} for general values of $v_2$ (giant magnon solutions 
were also constructed in~\cite{HST,Babichenko}). We must stress that in this case the periodicity condition cannot be imposed 
because the elliptic sine has infinite period and the string does not close. Also factors $\sqrt{(1-q^2) (\zeta_3 -\zeta_1)} / n \hbox{K} (\kappa )$ which had been canceled 
in the expressions for the angular momenta and windings do not cancel anymore. The third case corresponds to setting $\zeta_1=\zeta_2=\zeta_3$ and cannot be obtained 
unless we have equal angular frequencies, $\omega_1=\omega_2$.


\subsection{Solutions with pure NS-NS flux}

The cubic term in the polynomial $P_3(\zeta)$ is dressed with a factor $1-q^2$. Therefore in the case of pure NS-NS three-form flux 
the degree of the polynomial reduces to two and the solution can be written using trigonometric functions. In this limit 
\footnote{We can also take the limit directly in equation (\ref{etasn}) if we take into account that 
$\zeta_1$ goes to minus infinity when we set $q = 1$. In this limit the elliptic modulus vanishes but the factor $(1-q^2)\zeta_1$ 
in the argument of the elliptic sine remains finite and we just need to recall that $\hbox{sn} (x,0) = \sin x$.}
\be
\zeta^{\prime 2} = - 4 P_2 (\zeta) \ ,
\label{Qequation}
\ee
with $P_2(\zeta)$ the second order polynomial
\begin{align}
P_2 (\zeta) & = (\zeta - \omega_1^2) (\zeta - \omega_2^2) (\omega_1^2 - \omega_2^2) \bar{I}_1 + (\zeta - \omega_1^2)^2  v_2^2 \nonumber \\ 
& + (\zeta - \omega_2^2)^2 (v_1 + \omega_2)^2 = \omega^2 (\zeta - \tilde{\zeta}_1) (\zeta - \tilde{\zeta}_2) \ ,
\end{align}
where $\omega^2$ is 
\be
\omega^2 = (\omega_1^2 -\omega_2^2) \bar{I}_1 + (v_1 + \omega_2)^2 + v_2^2 \ .
\label{omega}
\ee
The solution to equation (\ref{Qequation}) is given by
\be
\zeta(\sigma) = \tilde{\zeta}_2 +( \tilde{\zeta}_1 - \tilde{\zeta}_2 ) \sin ^2 ( \omega \sigma ) \ ,
\ee
where we have set to zero an integration constant by performing a rotation. Therefore
\be
r_1^2(\sigma) = \frac{\tilde{\zeta}_2 - \omega_1^2}{\omega_2^2 - \omega_1^2} + \frac{\tilde{\zeta}_1 - \tilde{\zeta}_2}{\omega_2^2 - \omega_1^2} \, \sin^2
(\omega \sigma) \ .
\label{r1q1}
\ee
Periodicity of the radial coordinates implies that $\omega$ must be a half-integer number. 
\footnote{An important exception to this condition happens when $\omega = J + \bar{m}_2$, which is a solution even if it is not a half-integer. 
This value corresponds to the case of constant radii, where \cite{HN}
\[
\bar{I}_1 = -\left| \frac{2(v_1+\omega_2) v_2}{\omega_2^2-\omega_1^2} \right| \ , \quad 
J_1 = \frac{\bar{m}_2 J}{\bar{m}_2 - \bar{m}_1} \ , \quad J_2 = \frac{\bar{m}_1 J}{\bar{m}_1 - \bar{m}_2} \ , \quad E  = J - \bar{m}_1 \ .
\]
}
The relation between the winding numbers $\bar{m}_i$ and the constants $v_i$ and the frequencies $\omega_i$ is now rather simple. 
The periodicity condition for the angles implies
\begin{align}
\bar{m}_1 + \omega_2 & = \frac{(v_1 + \omega_2)(\omega_1^2 - \omega_2^2)}{\sqrt{(\omega_1^ 2-\tilde{\zeta}_1)(\omega_1^ 2 - \tilde{\zeta}_2)}}
= \frac{ \omega (v_1 + \omega_2)(\omega_1^2 - \omega_2^2)}{\sqrt{P_2(\omega_1^2)}} = \omega \, \hbox{sgn}(v_1 + \omega_2) \ ,\label{p1} \\
\bar{m}_2 + \omega_1& = \frac{v_2(\omega_1^2-\omega_2^2)}{\sqrt{(\omega_2^ 2-\tilde{\zeta}_1)(\omega_2^ 2 - \tilde{\zeta}_2)}}
= \frac{\omega v_2(\omega_1^2-\omega_2^2)}{\sqrt{P_2(\omega_2^2)}} = \omega \, \hbox{sgn} (v_2) \ . \label{p2}
\end{align}
From the definition of the angular momenta we find
\be
\frac{J_1}{\sqrt{\lambda}} = \omega_1-v_2 \ , \quad \frac{J_2}{\sqrt{\lambda}} = \bar{m}_1 - v_1 \ .
\label{p3}
\ee
We can now write the energy as a function of the winding numbers and the angular momenta. 
A convenient way to do this is recalling the relation between the energy and the Uhlenbeck constant. 
If we assume that both $v_1+\omega_2$ and $v_2$ are positive (the extension to the other possible signs of $v_1+\omega_2$ and $v_2$ is immediate)  
and we combine equations (\ref{H}) and (\ref{omega}) we can write
\be
E^2 = \lambda \big(  \omega^2 + \omega_1^2 - \omega_2^2 - 2v_1 \omega_2 - 2 v_2 \omega_1 \big) \ .
\ee
and thus using relations (\ref{p1})-(\ref{p3}) we conclude that
\be
E^2 = \lambda \bar{m}_1^2 + \big( 2 \sqrt{\lambda} J_1 - \lambda ( \omega - \bar{m}_2 ) \big) (\omega - \bar{m}_2) 
+ 2 \sqrt{\lambda} J_2 (\omega - \bar{m}_1) \ .
\label{EJ1J2}
\ee
Now we can use the Virasoro constraint (\ref{Virasoro2}) to write 
\be
J_1 = \frac{(J - \sqrt{\lambda} \omega)(\omega - \bar{m}_2)}{\bar{m}_1 - \bar{m}_2} \ , \quad J_2 = \frac{ J (\bar{m}_1 - \omega) 
+ \sqrt{\lambda} \omega ( \omega - \bar{m}_2 )}{\bar{m}_1 - \bar{m}_2} \ .
\ee
where $J = J_1 + J_2$ is the total angular momentum. Replacing these expressions in (\ref{EJ1J2}) we obtain the energy 
as a function of the winding numbers and the total momentum, 
\begin{equation}
E^2 = \lambda \big( \bar{m}_1^2 - \bar{m}_2^2 + 4 \omega \bar{m}_2- 3 \omega^2 \big) - 2\sqrt{\lambda} J (\bar{m}_1 + \bar{m}_2 -2\omega) \ .
\end{equation}


\section{Spinning strings in $AdS_3$}

In this section we will analyze the case where the strings are spinning in $AdS_3$. We will include no dynamics along $S^3$ so that the conserved charges 
of the solutions in this section will be the energy and spin. 
We will therefore choose the ansatz
\be
Y_3 + i Y_0 = z_0 (\sigma) \, e^{i \phi_0 (\tau, \sigma)} \ , \quad Y_1 + i Y_2 = z_1 (\sigma) \, e^{i \phi_1 (\tau, \sigma)} \ ,
\label{ansatzads}
\ee
where the angles are chosen as
\be
\phi_a (\tau, \sigma) = w_a \tau + \beta_a (\sigma) \ ,
\label{ansatzangleads}
\ee
with $a=0,1$. In this case periodicity requires
\be
z_a(\sigma + 2 \pi) = z_a (\sigma) \ , \quad \beta_a(\sigma + 2 \pi) = \beta_a(\sigma) + 2 \pi \bar{k}_a \ .
\ee
When we enter this ansatz in the world-sheet action in the conformal gauge we find 
\begin{equation}
L_{AdS_3} = \frac{\sqrt{\lambda}}{4\pi} \Big[ g^{ab} \left( z'_a z'_b + z_a z_a \beta_b'^2 - z_a z_a w_b^2 \right) 
-\frac {\tilde{\Lambda}}{2} \left( g^{ab} z_a z_b +1 \right) - 2 q z_1^2 ( w_0 \beta '_1 - w_1 \beta '_0 ) \Big] \ , \label{NRqads}
\end{equation}
where we have taken $g=\hbox{diag}(-1,1)$ and $\tilde{\Lambda}$ is the Lagrange multiplier needed to impose that 
the solutions lie on $AdS_3$, 
\be
z_1^2 - z_0^2 = - 1 \ . 
\label{adsthree}
\ee
The equations of motion for the radial coordinates are
\begin{align}
z''_0 & = z_0 \beta^{'2}_0 - z_0 w_0^2 - \tilde{\Lambda} z_0 \ , \label{z0prime} \\
z''_1 & = z_1 \beta^{'2}_1 - z_1 w_1^2 - \tilde{\Lambda} z_1 -2 q z_1 ( w_0 \beta'_1 - w_1 \beta'_0 ) \ , \label{z1prime}
\end{align}
and the equations for the angles are 
\be
\beta '_a = \frac{u_a + q z_1^2 \varepsilon_{ab} w_b}{g^{aa} z_a^2} \ , \\
\label{betaprime}
\ee
with $u_a$ some integrals of motion and $\varepsilon_{10}=+1$. The Virasoro constraints now read
\begin{align}
& z^{'2}_0 + z_0^2 (\beta ^{'2}_0 + w_0^2 ) = z^{'2}_1 + z_1^2 (\beta ^{'2}_1 + w_1^2) \ , \label{Virasoro1AdS} \\
& z_1^2 w_1 \beta '_1 - z_0^2 w_0 \beta '_0 =u_0 w_0 +u_1 w_1 =0 \ . \label{Virasoro2AdS}
\end{align}
The energy and the spin are given by
\begin{align}
E & = \sqrt{\lambda} \int_0^{2\pi} {\frac {d \sigma}{2 \pi} \left( z^2_0 w_0 - q z_1^2 \beta_1' \right)} \ , \label{EAdS} \\
S & = \sqrt{\lambda} \int_0^{2 \pi} {\frac {d \sigma}{2 \pi} \left( z^2_1 w_1 - q z_1^2 \beta_0' \right)} \label{S} \ .
\end{align}

As in the previous section, in order to construct general solutions for strings rotating in $AdS_3$ it will be convenient 
to introduce an analytical continuation of the ellipsoidal coordinates. The definition of this coordinate $\mu$ can be directly borrowed 
from the definition for the sphere with a change of sign,
\be
\frac{z_1^ 2}{\mu -w_1^ 2} - \frac{z_0^2}{\mu-w_0^2} = 0 \ .
\ee
If we order the frequencies such that $w_1 > w_0$ the range of the ellipsoidal coordinate will be $w_1^2 \leq \mu$. 
Now we can again make use of the Uhlenbeck constants to obtain a first order differential equation for this coordinate. 
In the case of the Neumann-Rosochatius system in $AdS_3$ the Uhlenbeck integrals 
satisfy the constraint $F_1 - F_0 = -1$, and thus we are again left with a single independent constant. 
To obtain the deformation of, say, $F_1$ by the NS-NS flux we can proceed in the same way as in section 2. 
After some immediate algebra we conclude that
\be
\bar{F}_1=z_1^2 (1-q^2) +\frac{1}{w_1^2 -w_0^2} \left[ (z_1 z'_0-z'_1 z_0)^2 +\frac{(u_0+q w_1)^2}{z_0^2} z_1^2 +\frac{u_1^2}{z_1^2} z_0^2 \right] \ .
\ee
The Hamiltonian can also be written now using the deformed Uhlenbeck constants and the integrals of motion $u_a$,
\be
H = \frac {1}{2} \sum_{a=0}^1 \big[ g_{aa} w_a^2 \bar{F}_a - u_a^2 \big] + q u_1 w_0 \ .
\label{HAdS}
\ee
Now we need to note that
\be
(z_1 z'_0 -z_0 z'_1)^2=\frac{\mu^{\prime 2}}{4 (\mu -w_1^2) (\mu-w_0^2)} \ . 
\ee
When we solve for $\mu'^2$ in the deformed integral we find that
\be
\mu'^2 = -4 Q_3 (\mu) \ ,
\label{diffeqzmu}
\ee
where $Q_3(\mu)$ is the third order polynomial, 
\begin{align}
Q_3 (\mu)
& = (1-q^2) (\mu -w_1^2 )^2 (\mu -w_0^2) + (\mu -w_1^2 ) (\mu -w_0^2) (w_0^2 -w_1^2) \bar{F}_1 \notag \\
& + (\mu - w_1^2 )^2  (u_0+q w_1)^2 + (\mu -w_0^2)^2 u_1^2 = (1-q^2 ) \prod_{i=1}^3 (\mu -\mu_i)  \ .
\end{align}
This equation is an analytic continuation of the spherical one, so we can write
\be
z_0^2(\sigma) = \frac{\mu_3 - w_0^2}{w_1^2 -w_0^2} +\frac{\mu_2 -\mu_3}{w_1^2 -w_0^2} \, \hbox{sn}^2 
\big( \sigma \sqrt{(1-q^2) (\mu_3 -\mu_1)} , \nu \big) \ ,
\label{z0elliptic}
\ee
where the elliptic modulus is $\nu = (\mu_3 - \mu_2)/(\mu_3 - \mu_1)$. 
As in the case of strings rotating in $S^3$ we must perform now an analysis of the roots of the polynomial. We need to choose $\mu_3 > \mu_1$ 
to make sure that the argument of the elliptic sine is real, and $\mu_3 > \mu_2$ to have $\nu>0$, together with $\mu_2>\mu_1$ to keep $\nu<1$. 
Furthermore the $AdS_3$ condition (\ref{adsthree}) implies that $z_0^2 \geq 1$ which constrains $\mu_2$ and $\mu_3$ to be greater or equal than $w_1^2$. This restriction does 
not apply to $\mu_1$. Note that this hierarchy of roots implies that not all possible combinations of the parameters $u_i$, $w_i$ and $\bar{F}_1$ are allowed. 
The periodicity condition on the radial coordinates now implies that~\footnote{Again there are four different cases where this condition 
must be modified. When either $u_0+q w_1=0$ or $u_1=0$ the periodicity condition becomes 
$\frac{\pi}{2} \sqrt{(1-q^2) (\mu_3 -\mu_1)} = n' \, \hbox{K} (\nu)$ because of changes of branch when we take the square root of (\ref{z0elliptic}). The other two cases 
are the degenerate limits where there is no periodicity.}
\be
\pi \sqrt{(1-q^2) (\mu_3 -\mu_1)} = n' \hbox{K} (\nu) \ ,
\ee
with $n'$ an integer number. From the periodicity condition on $\beta_1$,
\be
2 \pi \bar{k}_1 = \int_0^{2\pi}{\beta'_1 d\sigma}=\int_0^{2\pi}{\left( \frac{u_1}{z_1^2} + q w_0  \right) d\sigma} \ ,
\ee
we can write
\be
\frac{\bar{k}_1 - q w_0}{u_1}= \int_0^{2\pi} \frac{d\sigma}{2\pi} \frac{1}{z_1^2} \ .
\ee
Performing the integration we find
\be
\bar{k}_1 - q w_0= \frac{u_1 (w_1^2 -w_0^2)}{(\mu_3 - w_1^2) 
\hbox{K}  (\nu) } \Pi \left( \frac{\mu_3 -\mu_2}{\mu_3 - w_1^2}, \nu \right) \ .
\ee
The periodicity condition for $\beta_0$ implies that
\be
2 \pi \bar{k}_0 = \int_0^{2\pi}{\beta'_0 d\sigma} = \int_0^{2\pi}{\left( -\frac{u_0}{z_0^2} + q w_1 \frac{z_1^2}{z_0^2} \right) d \sigma} \ .
\label{beta0periodicity}
\ee
Now we must recall that we are working in $AdS_3$ instead of its universal covering. The time coordinate should therefore be single-valued, 
and thus we have to exclude windings along the time direction. When we set $\bar{k}_0=0$ equation (\ref{beta0periodicity}) becomes
\be
\frac{q w_1}{u_0 +q w_1}= \int_0^{2\pi} \frac{d\sigma}{2\pi} \frac{1}{z_0^2} \ ,
\ee
that we can integrate to get
\be
q w_1= \frac{(u_0 +q w_1) (w_1^2-w_0^2 )}{(\mu_3 - w_0^2) \hbox{K} (\nu)} \Pi \left( \frac{\mu_3 -\mu_2}{\mu_3 - w_0^2} , \nu \right)  \ .
\ee
In the same way we can perform an identical computation to obtain the energy and the spin. From equation (\ref{EAdS}) we get
\be
\frac{E} {\sqrt{\lambda}} + q u_1 - q^2 w_0= (1-q^2) w_0 \int_0^{2\pi}{\frac{d\sigma}{2\pi} z_0^2} \ ,
\ee
and thus
\be
\frac{E} {\sqrt{\lambda}} = q^2 w_0- q u_1 +\frac{(1-q^2) w_0}{w_1^2 -w_0^2} \left[ \mu_3 - w_0^2 -(\mu_3 - \mu_1) 
\left( 1-\frac{ \hbox{E} (\nu)}{\hbox{K} (\nu)} \right) \right] \ .
\ee
Repeating the same steps with (\ref{S}) we obtain an expression for the spin,
\be
\frac{S} {\sqrt{\lambda}} - q u_0 = (1-q^2) w_1 \int_0^{2\pi}{\frac{d\sigma}{2\pi} z_1^2} \ ,
\ee
and thus
\be
\frac{S} {\sqrt{\lambda}} =q u_0 +\frac{(1-q^2) w_1}{w_1^2 -w_0^2} \left[ \mu_3 - w_1^2 - (\mu_3 - \mu_1) 
\left( 1-\frac{ \hbox{E} (\nu)}{\hbox{K} (\nu)} \right) \right] \ .
\ee
These expressions for the energy and the spin can be used to rewrite the first Virasoro constrain (\ref{Virasoro2AdS}) as
\be
w_1 E - w_0 S =\sqrt{\lambda} w_0 w_1 \ .
\ee
which is already a very closed expression. But what we want is a relation involving only $E$, $S$ and~$\bar{k}_1$. 
As in the case of strings spinning in $S^3$, obtaining this relation for arbitrary values of $q$ is again a lengthy and complicated problem and we will not discuss it here. We will consider instead in the following subsection the case where a simple treatment is possible which is the limit of pure NS-NS flux of the above solutions. 

To conclude our discussion we are going to briefly discuss the choices of parameters that make the discriminant of $Q_3(\mu)$ equal to zero. 
As in the previous section, there are only three possible cases where the discriminant vanishes. 
The first is the constant radii case, where $\mu_2=\mu_3$. However this limit is not always well defined. The second case corresponds to the limit $\kappa=1$ and 
it is obtained when $\mu_1=\mu_2$. In this case there is no periodicity condition because the elliptic sine has infinite period and thus the string does not close. 
Also the factors $\sqrt{(1-q^2) (\mu_3 -\mu_1)}/n \hbox{K} (\nu )$ do not cancel anymore in the expressions for the energy, the spin and the winding number. 
The third case corresponds to $\mu_1=\mu_2=\mu_3$ and requires setting $w_0=w_1$.


\subsection{Solutions with pure NS-NS flux}

As in the case of strings rotating in $S^3$, in the limit of pure NS-NS three-form flux the above solutions can be written 
in terms of trigonometric functions. Now (\ref{diffeqzmu}) reduces to
\be
\mu^{\prime 2} = - 4 Q_2 (\mu) \ ,
\label{Q2equation}
\ee
with $Q_2(\mu)$ the second order polynomial
\begin{align}
Q_2 (\mu) & = (\mu - w_1^2) (\mu - w_0^2) (w_0^2 - w_1^2) \bar{F}_1 + (\mu - w_0^2)^2  u_1^2 \nonumber \\ 
& + (\mu - w_1^2)^2 (u_0 + w_1)^2 = \omega'^2 (\mu - \tilde{\mu}_1) (\mu - \tilde{\mu}_2) \ ,
\end{align}
where $\omega'^2$ is 
\be
\omega'^2 = (w_0^2-w_1^2) \bar{F}_1 +(u_0+w_1)^2 +u_1^2 \ .
\label{omegaprime} 
\ee
Thus we conclude that
\be
z_0^2(\sigma) = \frac{\tilde{\mu}_2 - w_0^2}{w_1^2 - w_0^2} + \frac{\tilde{\mu}_1 - \tilde{\mu}_2}{w_1^2 - w_0^2} \, \sin ^2 (\omega' \sigma ) \ .
\label{z0trig}
\ee
The periodicity condition on the radial coordinates implies now that $\omega'$ should be a half-integer number. 
The frequencies $w_a$ and the integration constants $u_a$ are related to the energy, the spin and the winding number $\bar{k}_1$ by 
\begin{align}
w_1 & = \omega' \, \hbox{sgn} \, (u_0+w_1)\ , \quad  \omega' = (\bar{k}_1-w_0) \, \hbox{sgn}(u_1) \label{q1} \ , \\
S & = \sqrt{\lambda} u_0 \ , \quad E = \sqrt{\lambda} (w_0 -u_1)=\frac{w_0}{w_1} S+\sqrt{\lambda} w_0 \label{q2} \ .
\end{align}
Recalling now the Virasoro condition (\ref{Virasoro1AdS}) the spin can be written as
\be
S =  \sqrt{\lambda} \, \frac{(\bar{k}_1-\omega')^2 \omega'}{2 \bar{k}_1 (2 \omega' - \bar{k}_1)} \ , 
\ee
while the energy is given by
\be
E = \sqrt{\lambda} \, \frac{\bar{k}_1^3 - 3 \bar{k}_1^2 \omega' + \bar{k}_1 \omega^{\prime 2} 
+ \omega^{\prime 3}}{2 \bar{k}_1 (\bar{k}_1 - 2 \omega')} \ .
\ee
We must note that we still have to impose a restriction on the parameters.  
This restriction comes from imposing that the discriminant of $Q_2 (\mu)$ must be positive 
and taking the region in the parameter space with the correct hierarchy of roots. 
This condition can be written as
\be
 | 2 (u_0 + w_1) u_1 | \leq \left| \bar{F}_1 (w_1^2 -w_0^2)\right| =\left| \omega ^{\prime 2} -(u_0 +w_1)^2-u_1^2 \right| \ .
\ee
The inequality is saturated in the cases of constant radii.


\section{Concluding remarks}

In this article we have found a general class of solutions of the flux-deformed Neumann-Rosochatius system. The solutions 
that we have constructed correspond to closed strings with non-constant radii rotating in $AdS_3 \times S^3 \times T^4$ with mixed R-R 
and NS-NS three-form fluxes. We have considered the cases where the string is rotating either in $S^3$ with two different angular momenta 
or in $AdS_3$ with one spin. The corresponding solutions can be expressed in terms of Jacobian elliptic functions. In the limit 
of pure NS-NS flux the elliptic functions reduce to trigonometric functions. This reduction of the problem allows to write rather compact 
expressions for the classical energy of the spinning string as a function of the corresponding conserved quantities. 

The simplification in the limit of pure NS-NS flux is an appealing result already present in the case of the constant radii solutions studied in~\cite{HN}. 
For the solutions that we have constructed in this article the reduction implied by the presence of pure NS-NS flux appears as a consequence 
of the degeneration of the elliptic curve governing the dynamics of the problem. Furthermore, in this limit the theory can be described 
by a WZW model. It would be quite interesting to explore this limit in more detail and the relation of our approach to the solution of the WZW model. 
An important issue in this direction concerns the fate of the scattering matrix of the problem in the case of pure NS-NS flux. 
A complementary question for a better understanding of the quantum corrections to the spinning strings should come from the analysis of the spectrum 
of small quadratic fluctuations around the circular solutions. 

Another interesting question is the extension of our analysis to other possible deformations of the backgrounds of type IIB string theory. 
An engaging case is that of the $\eta$-deformation of the $AdS_5 \times S^5$ background~\cite{DMV}. This is a much more complicated 
problem than the deformation by flux that we have studied in here, because the deformation comes from the breaking of the isometries 
of the metric down to the Cartan algebra. However closed strings rotating in $\eta$-deformed $AdS_5 \times S^5$ have been shown recently 
to lead to an integrable extension of the Neumann-Rosochatius system~\cite{AM}. It would be interesting to continue our analysis in this article 
to find general solutions of this $\eta$-deformed Neumann-Rosochatius system. 


\vspace{8mm}

\centerline{\bf Acknowledgments}

\vspace{2mm}

\no
The work of R.~H. is supported by MICINN through a Ram\'on y Cajal contract and grant FPA2011-24568, 
and by BSCH-UCM through grant GR58/08-910770. J.~M.~N. wishes to thank the Instituto de F\'{\i}sica Te\'orica UAM-CSIC 
for kind hospitality during this work. 


\appendix

\renewcommand{\theequation}{\thesection.\arabic{equation}}
\csname @addtoreset\endcsname{equation}{section}

\section{Spinning strings in $AdS_3 \times S^3$ with pure NS-NS flux}

In this appendix we will consider the case where the string is allowed to rotate both in $AdS_3$ and $S^3$. We will restrict the analysis 
to the limit of pure NS-NS flux. 
As the pieces in the lagrangian describing motion in $AdS_3$ and $S^3$ are decoupled the equations 
for the corresponding coordinates are directly given by (\ref{r1prime})-(\ref{alphaprime}) and (\ref{z0prime})-(\ref{betaprime}). 
The coupling between the $AdS_3$ and $S^3$ factors comes from the Virasoro constraints, 
\begin{align}
& z^{'2}_0 + z_0^2 (\beta ^{'2}_0 + w_0^2 ) = z^{'2}_1 + z_1^2 (\beta ^{'2}_1 + w_1^2)
+ \sum_{i=1}^2 \big( r^{'2}_i + r_i^2 (\alpha^{'2}_i + \omega^2_i ) \big)  \ , \\
& z_1^2 w_1 \beta '_1 + \sum_{i=1}^2 r_i^2 \omega _i \alpha '_i  = z_0^2 w_0 \beta '_0 \ .
\end{align}
The second Virasoro constrain can be rewritten as
\be
\omega_2 J_1 +\omega_1 J_2 +w_1 E -w_0 S= \sqrt{\lambda} \left( \omega_1 \omega_2 + w_0 w_1 +q \omega_1 \bar{m_1} \right) \ .
\ee
With these relations at hand, together with equations (\ref{p1}), (\ref{p2}), (\ref{q1}) and (\ref{q2}), it is immediate to write 
the angular momenta and the energy as functions of $\omega$, $\omega'$, the winding numbers $\bar{m}_1$, $\bar{m}_2$ and $\bar{k}_1$, 
and the spin $S$ and the total angular momentum $J$. In the case where $w_0+\bar{k}_1=-w_1=-\omega'$ we conclude that
\begin{eqnarray}
J_1 \!\! & = & \!\! \big[ -\bar{k}_1^2 (\sqrt{\lambda} \omega ' + 2 S) + 2 \bar{k}_1\big( \sqrt{\lambda} \omega^{\prime 2} + 2 \omega' S 
+ (\bar{m}_2 - \omega) (\sqrt{\lambda} \omega - J) \big) \\
& + & \!\! \omega' \big( \sqrt{\lambda} (\bar{m}_1^2 - \bar{m}_2^2 - \omega^{\prime 2} + \omega^2 )
- 2 (\bar{m}_1 - \bar{m}_2) J \big) \big] / 
\big( 2 (\bar{m}_1 - \bar{m}_2) (\bar{k}-2\omega') \big) \ , \nonumber \\
J_2 \!\! & = & \!\! \big[ \, \bar{k}_1^2 (\sqrt{\lambda} \omega'+2S) - 2 \bar{k}_1\big( \sqrt{\lambda} \omega^{\prime 2} + 2 \omega' S - \bar{m}_1 J 
+ \omega (\sqrt{\lambda} \bar{m}_2 - \sqrt{\lambda} \omega + J) \big) \\
& - & \!\! \omega' \big( \sqrt{\lambda} (\bar{m}_1^2 - \bar{m}_2^2 - \omega^{\prime 2} + \omega^2) 
+ 2 (\bar{m}_1 - \bar{m}_2) J \big) \big] / \big( 2 (\bar{m}_1 - \bar{m}_2) 
(\bar{k}_1-2\omega') \big) \ , \nonumber \\
E \!\! & = & \!\! \big[ \sqrt{\lambda} \big( \bar{k}_1^2 + \bar{m}_1^2 - (\bar{m}_2 -3\omega)(\bar{m}_2 -\omega) \big)-2 \bar{k}_1 (2 \sqrt{\lambda} \omega' +S) \\
& + & \!\! \omega' (3 \sqrt{\lambda} \omega' +4S) -2 J (\bar{m}_1 + \bar{m}_2 -2\omega) \big] / \big( 2 (\bar{k}_1-2 \omega') \big) \ . \nonumber
\end{eqnarray}
If we choose $w_0 + \bar{k}_1 = w_1 = \omega'$ we find
\begin{eqnarray}
J_1 \!\! & = & \!\!\ \big[ -\bar{k}_1^2 (\sqrt{\lambda} \omega ' + 2 S) + 2 \bar{k}_1 (-\sqrt{\lambda} \omega^{\prime 2} -2 \omega' S 
+ (\bar{m}_2 - \omega) (\sqrt{\lambda} \omega - J)) \\ 
& + & \!\! \omega' \big( \sqrt{\lambda} (\bar{m}_1^2 -\bar{m}_2^2 - \omega^{\prime 2} +4 \bar{m}_2 \omega -3 \omega^2 )
- 2 (\bar{m}_1 + \bar{m}_2 -2 \omega \big) J) \big] / 
\big( 2 \bar{k}_1 (\bar{m}_1 - \bar{m}_2) \big) \ , \nonumber \\
J_2 \!\! & = & \!\! \big[ \, \bar{k}_1^2 (\sqrt{\lambda} \omega'+2S) + 2 \bar{k}_1 \big( \sqrt{\lambda} \omega^{\prime 2} +2 \omega' S + \bar{m}_1 J 
- \omega (\bar{m}_2 - \omega + J) \big) \\
& - & \!\! \omega' \big( \sqrt{\lambda} (\bar{m}_1^2 - \bar{m}_2^2 - \omega^{\prime 2} + 4 \bar{m}_2 \omega -3 \omega^2)
- 2 (\bar{m}_1 + \bar{m}_2 -2\omega ) J\big) \big] / 
\big( 2 \bar{k}_1 (\bar{m}_1 - \bar{m}_2) \big) \ , \nonumber \\
E \!\! & = & \!\! \big[ \sqrt{\lambda} \big( \bar{k}_1^2 + \bar{m}_1^2-(\bar{m}_2 -3\omega)(\bar{m}_2 -\omega) - \omega'^2 \big) -2 \bar{k}_1 S \\
& - & \!\! 2 J (\bar{m}_1 + \bar{m}_2 -2\omega) \big] / \big( 2 \bar{k}_1 \big) \ . \nonumber
\end{eqnarray}
When we take the limit $\bar{k}_1 \rightarrow 0$, $S \rightarrow 0$ and $\sqrt{\lambda} \omega' \rightarrow E$ we recover the expressions 
from section 3 in both cases. In a similar way when we set to zero the angular momenta, the winding numbers and $\omega$ we recover the analysis in section 4. 
We can also reproduce the solutions of constant radii analyzed in~\cite{HN}. In this case, when $w_0+\bar{k}_1=-w_1$ the angular momenta are given by
\be
J_1 =\frac{\bar{k}_1 S + \bar{m}_2 J}{\bar{m}_2 - \bar{m}_1} \ , \quad J_2 = \frac{\bar{k}_1 S + \bar{m}_1 J}{\bar{m}_1 - \bar{m}_2} \ , 
\ee
and the energy reduces to 
\be
E = - S \pm (J- \sqrt{\lambda} \bar{m}_1) \ .
\ee
In the case where $w_0+\bar{k}_1=w_1$ the angular momenta are
\be
J_1 = \frac{\bar{k}_1 S + \bar{m}_2 J}{\bar{m}_2 - \bar{m}_1} \ , \quad J_2 = \frac{\bar{k}_1 S + \bar{m}_1 J}{\bar{m}_1 - \bar{m}_2} \ , 
\ee
and the energy becomes 
\be
E = S \pm \sqrt{(J - \lambda \bar{m}_1)^2-4 \sqrt{\lambda} \bar{k}_1 S} \ .
\ee


\end{document}